\documentclass{elsart}
\usepackage{graphicx,amssymb} 
\journal{Elsevier Preprint} 
\begin{document} 
\begin{frontmatter} 
\title{A prototype for the AMS-RICH experiment} 
\author[gaelle]{{G. Boudoul}\thanksref{now}}
%\corauth[cor]{Corresponding author.} 
%\ead{boudoul@isn.in2p3.fr} 
\address[gaelle]{Institut des Sciences Nucl\'eaires (CNRS/IN2P3 - UJF), 53 Avenue des Martyrs,
38026 Grenoble cedex, France.\\
boudoul@isn.in2p3.fr} 
%\author[AMS]{AMS-RICH Collaboration\thanksref{now}}
\thanks[now]{For the AMS-RICH 
Collaboration: INFN Bologna, ISN Grenoble, LIP Lisbon, CIEMAT Madrid, U. Maryland, and 
UNAM Mexico}
\begin{abstract} 
The AMS spectrometer will be installed on the International Space Station in
2005. Among other improvements over the first version of the
instrument, a ring imaging Cherenkov detector (RICH) will be added and
should open a new window for cosmic-ray physics, allowing isotope separation up
to A$\approx$25 between 1 and 10 GeV/c and element identification up to
Z$\approx$25 between threshold and 1 TeV/c/nucleon. It should also contribute
to the high level of redundancy required for AMS and reject efficiency albedo
particles. A second generation prototype has been operated for a few months :
the architecture and the first results are presented.

\end{abstract}
%\begin{keyword} 
%\end{keyword} 
\end{frontmatter}
\section{Introduction} 
The AMS spectrometer \cite{barrau} will be implemented on the 
International Space Station in 2005. The instrument will 
be made of a superconducting magnet which inner volume will be mapped with a tracker 
consisting of 8 planes of silicon microstrips with a set of detectors for
particle identification placed above and below the magnet: scintillator hodoscopes, electromagnetic calorimeter (ECAL), 
transition radiation detector (TRD) and ring imaging Cherenkov (RICH). This
contribution 
is devoted to a study a second generation prototype aiming at the RICH testing. \\

\begin{figure}[h]  
%\vspace{-0.3cm}
\begin{center}             % fig 1
%\vspace*{2.0mm} % just in case for shifting the figure slightly down
\includegraphics[scale=.3]{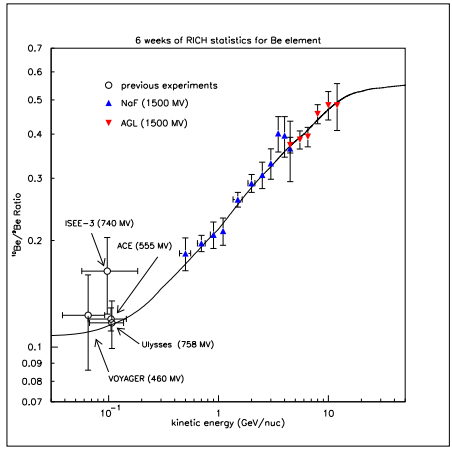} 
\caption{\label{fig:alexi}
Expected statistics for the $^{10}$Be in 6 weeks of counting 
with AMS \cite{alexi}.}
\end{center}
\vspace{-0.2cm}
\end{figure}

The physics capability of the RICH counter has been investigated 
by simulations \cite{buenerd}. It should provide unique informations among the AMS 
detectors by several respects :
\begin{itemize}
\item Isotopes separation up to A$\approx$25 at best, over a momentum range extending
from about 1-2 GeV/c up to around 13 GeV/c.
\item Identification of chemical elements up to Z$\approx$25 at best, up to approximately
1 TeV/nucleon.
\item High efficiency rejection of albedo particles for momenta above the
threshold, between 1 GeV/c and 3.5 GeV/c depending on the type of
radiator.
\end{itemize}
The RICH counter will allow to collect a unique sample of nuclear astrophysics
data with unprecedented statistical significance over a momentum range totally
unexplored for the most interesting isotopes. Fig. \ref{fig:alexi} shows, as an example, the $^{10}$Be to $^{9}$Be ratio with 6 weeks of 
counting time \cite{alexi}. Both the number of events and the covered energy range will 
dramatically improve the available data (lower left points on the plot).
 
Recent works \cite{donato} have emphasized the importance 
of measuring cosmic nuclei spectra for: 1) Setting strong constraints on the astrophysical 
and cosmic ray propagation parameters of the galaxy : the diffusion coefficient 
normalisation and its spectral index, the halo thickness, the Alfv\'en velocity and the 
convection velocity; 2) Increasing the sensitivity to new physics search for supersymmetric 
particles or primordial back holes; 3) Testing for the nature of the cosmic-ray sources :
 supernovae, stellar flares, Wolf-Rayet stars, etc ...

\section{New prototype}

The second generation prototype has been developed by the RICH group of the AMS collaboration. It is made of one half module of the 
final counter and has been operated for a few months.

It is equipped with R7600-M16 PMTs from Hamamatsu Inc. Ninety-six units are
used in the prototype, providing 1526 pixels. The R7600-M16 is a 16 pixels PMT 
($16\,\times\,(4.5\times4.5~mm^2$)) with 
12 metal channel dynodes and a borosilicate glass window. The high voltage divider used 
is a compromise between single photoelectron resolution ($\frac{\sigma}{Q} \approx 0.5$) and 
linearity. Its total resistivity was fixed at 80 M$\Omega$, which allows a very low
power consumption and remains compatible with the expected trigger rate
around 1kHz. The front-end electronics  is placed 
next to the PMT on a flex connector linked to the readout bus.
Each PMT is equipped with solid light guides to collect the Cherenkov photons and to 
reduce the dead-space between photocathodes.
This prototype is installed in the same instrumental setup as the previous
version \cite{thuillier} and uses the same trigger and tracker system, which includes three NE102 
plastic scintillator units, coupled with RTC-2262B PMTs, for 
trigger definition, and a set of three X-Y multiwire proportional chambers of $320\times320\,\mbox{mm}^2$
with 2 mm wire step, and delay lines readout.\\
In cosmic-ray tests, the 3 available space points 
  provide the trajectory reconstruction in the radiator 
with a $\sigma_{track}\approx1.9\,\mbox{mm}$ in both directions, by means of $\chi^2$
optimization procedure.
\begin{figure}
\begin{center}
\includegraphics[scale=.35]{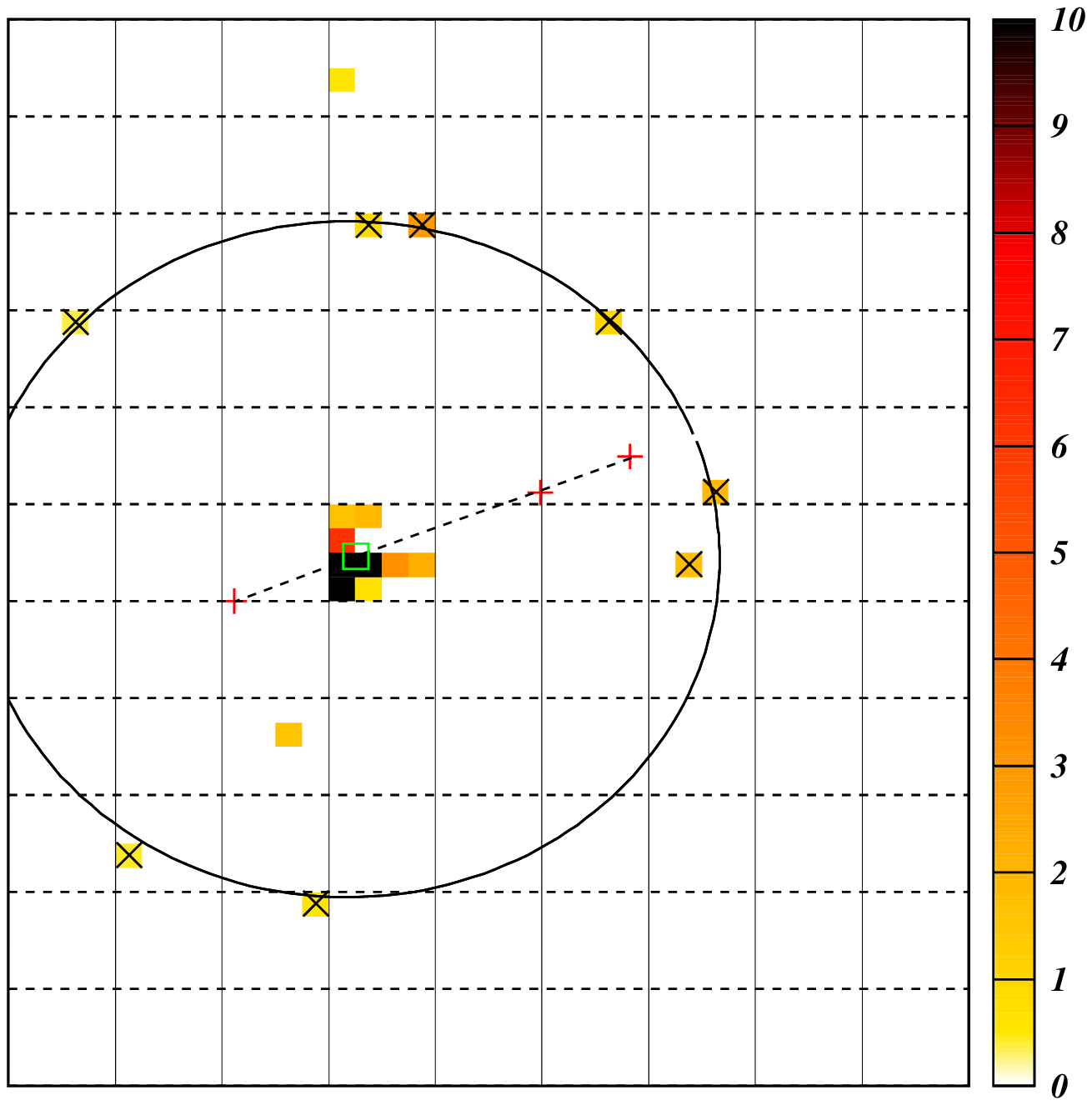}
\includegraphics[scale=.35]{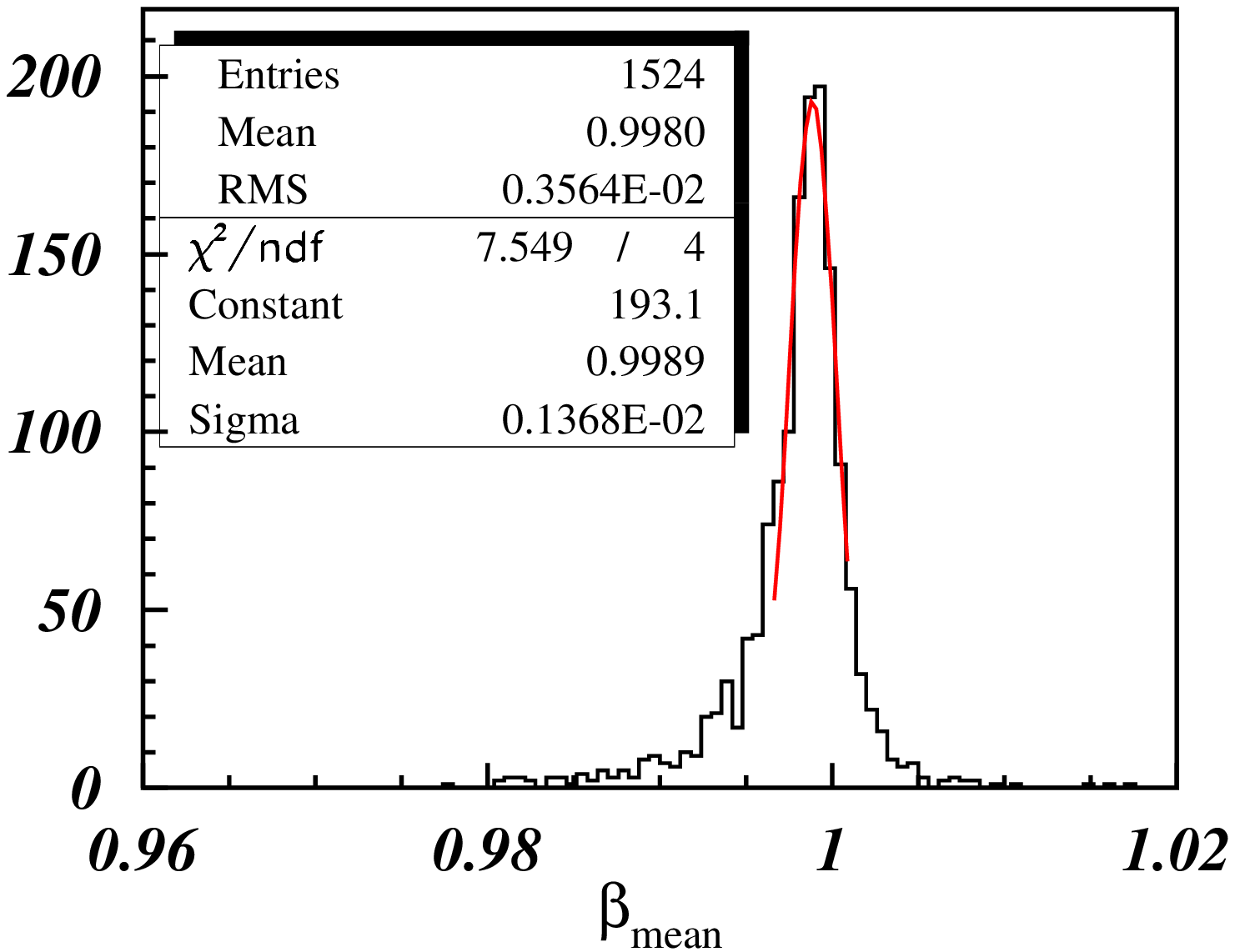}
\caption{\label{fig:proto}Left: Cherenkov ring produced by a cosmic ray. Note
the direct PMT hit (square at the center of the circle) on the reconstucted
trajectory (dotted line). Right: Reconstructed
velocity for n=1.03}
\end{center}
%\vspace{-1.cm}
\end{figure}

The main goals of this new generation of prototype are to validate the complex assembly 
procedure, check the readout electronics settings and DAQ procedure for all the 
output channels, investigate the PMT+electronics response dynamics, measure the 
counter velocity resolution, test the whole structure when submitted to 
vibrations, and validate the magnetic shielding efficiency.

It will also be tested with a secondary ion beam at CERN using a fragmentation target. The
sample of nuclei from H to Fe should allow an accurate check of the charge measurement resolution of the
counter.
\section{First results with cosmic rays}
Fig \ref{fig:proto} left gives an example of a Cherenkov ring obtained during
a cosmic ray run.\\
The best experimental velocity resolution obtained for cosmic ray test was
$\frac{\delta\beta}{\beta}=15\mbox{ }10^{-3}$ with $\overline{n} = 1.33$ NaF radiator 
(Cherenkov threshold around 480 MeV/nucleon), 
$\frac{\delta\beta}{\beta}=2.5\mbox{ }10^{-3}$ with $\overline{n} = 1.03$
aerogel radiator as shown on Fig \ref{fig:proto} right, and $\frac{\delta\beta}{\beta}=3.5\mbox{ }10^{-3}$ with
$\overline{n} = 1.05$
(Cherenkov threshold around 3.5 GeV/nucleon). These resolutions are limited by the 
multiwire chambers track reconstruction accuracy. Those results will be improved by
increasing the distance between chambers. Both the electronics and the PMTs
behave as expected.\\
The study has shown a good agreement between data and simulation and give
confidence in the expected performances of the final counter.

%\begin{figure}
%\vspace{-1.cm}
%\begin{center}
%\includegraphics[scale=.3]{beta.ps}
%\vspace{-1.7cm}
%\caption{\label{CR} Reconstructed spectrum for n=1.03. Data (points) and
%MC simulation (histogram) are in a very
%good agreement}
%\end{center}
%\end{figure}

\end{document}